\documentclass[%
reprint,
 amsmath,amssymb,
 aps,
]{revtex4-2}

\usepackage[dvipdfmx]{graphicx}
\usepackage{color}
\usepackage{bm}
\usepackage{amsmath,amssymb}
\usepackage{braket}

\begin{document}


\title{Selective excitation of a single rare-earth ion in an optical fiber}

\author{Kaito Shimizu$^{1}$}
\email{E-mail address: 1223702@ed.tus.ac.jp}

\author{Kazutaka Katsumata$^{1}$}

\author{Ayumu Rikuta$^{1}$}

\author{Tsuyoshi Kanemoto$^{1}$}

\author{Kei Sakai$^{1}$}

\author{Tomo Osada$^{1}$}

\author{Kaoru Sanaka$^{1}$}

\affiliation{$^{1}$ Department of Physics, Tokyo University of Science, Shinjuku, Tokyo 162-8601, Japan}

\date{\today}

\begin{abstract}
Fiber-coupled single-photon source is an essential component for the implementation of optical quantum communication technologies. Using the rare-earth ion doped in an optical fiber as an emitter is a significant method to construct such photon source at room temperature, as well as achieving high coupling and channeling efficiency. In this study, we experimentally demonstrated the generation of single photons at room temperature by selectively exciting a sole rare-earth ion isolated within a tapered silica fiber. The key advantages of our method are the ability to manipulate a purely single ion, and the efficient collection of photons from the guided mode of the fiber, owing to the single ion's emission of photons directly within the fiber. These features make our system a promising building block for realizing all-fiber-integrated optical quantum networks. We have also measured the optical lifetime of a single neodymium ion in the tapered fiber, and the result supports that the single-photon correlation time is practically determined by the absorption time of the ion. 
\end{abstract}

\maketitle

\clearpage
\section{Introduction}
The development of efficient and reliable single-photon sources has been desired to realize various optical quantum information technologies such as quantum computing~\cite{quantumcomputing01}, quantum random number generation~\cite{random01, random02} and quantum imaging~\cite{imaging01, imaging02}.
In particular, fiber-coupled single-photon source is an essential component for realizing quantum internet~\cite{internet01, internet02, internet03} and quantum communication tasks such as quantum key distribution~\cite{qkd01, qkd02, qkd03}.
The essence of such single-photon source is to achieve high coupling efficiency between the tapered fiber and the emitter. The emitter can vary from semiconductor quantum dots~\cite{qdot01, qdot02, qdot03}, single molecules~\cite{molecule01}, hexagonal boron
nitride(hBN)~\cite{hbn01} to nanodiamonds including the Nitrogen-Vacancy(NV) centers~\cite{nanodiamond01}.
Among these emitters, rare-earth(RE) ions have attracted much attention~\cite{singlerare01, singlerare02, singlerare03} due to their stable luminescence of the 4f transitions, and their ability to operate at room temperature.

In this paper, we demonstrate a novel experimental method to generate single photons by exciting a sole RE ion doped within a tapered optical fiber. This method builds up on our previously established scheme where spatially isolated RE ions are obtained by tapering a RE ion doped silica fiber~\cite{Yb01, Nd01}, as well as improving some of the weakness of our previous work: the fundamentally limited collection efficiency of the photons and the lack of extendibility of the system. The main concept to aim such improvement is to excite the RE ion “selectively” by focusing the pump laser onto a single isolated ion, and collect the single photons from the guided mode of the fiber. Using this system, we show that single photons are obtained at room temperature, proved by the autocorrelation measurement. We also achieved to measure the optical lifetime of a single neodymium ion(Nd$^{3+}$) using our ion doped fiber, which is a novel result to the best of our knowledge. This result was used to estimate its influence on the single-photon correlation time.

 \begin{figure}[htp]
 \centering
\includegraphics[width=8cm]{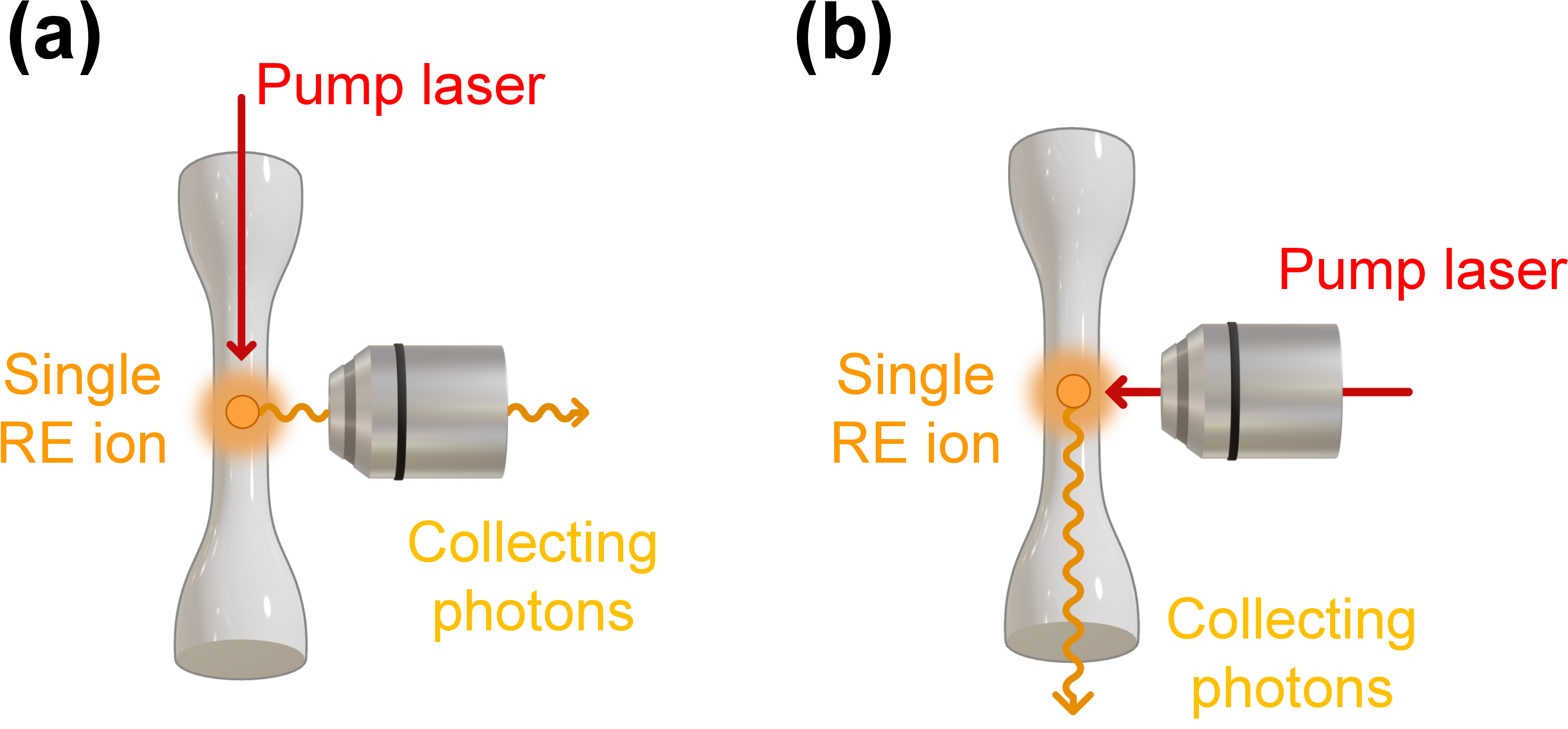}
\caption{\label{fig.selective}
(a) Method to excite single RE ions in the tapered fiber we previously proposed: the pump light is injected from the direction of the guided mode, and the photons are collected by an objective lens from the side of the fiber.
(b) Method to excite a sole single RE ion within the tapered fiber we propose in this paper: the pump light is injected from the side of the fiber, and photons are collected by coupling them to the guided mode. 
}
 \end{figure}

Figure \ref{fig.selective} compares the two schemes of our RE ion doped fiber system. For both schemes, we prepare a RE ion doped silica fiber which undergoes a tapering process, enabling access to single ions spatially separated beyond diffraction limit. Figure \ref{fig.selective}(a) excites the ensemble of ions by coupling the pump laser into the fiber guided mode. Emission from a single RE ion is resolved and collected by using an objective lens placed at the side of the fiber~\cite{Yb01, Nd01}. Figure \ref{fig.selective}(b), the new scheme we introduce in this work as the “selective excitation” method, instead excites a single RE ion by focusing the pump laser onto one of the spatially resolved ions. Photons emitted from the ion within the fiber which couple and propagate into the fiber guided mode are collected. The difference between these two schemes is not merely a swap in the direction of the excitation and photon collection. The collection efficiency of the former scheme is fundamentally limited by the numerical aperture of the objective lens. With our experimental setup, the proportion of the photons that can be captured by the objective lens among all the photons emitted by the ion is at most 6.7\% since the numerical aperture of the oblective lens used in our experiment is 0.5. On the other hand, using the latter scheme, one can theoretically achieve a collection efficiency of 31\% when photons are collected from both sides of the fiber (see Appendix A for the details of the calculation).

In addition, we can expect further technical extensions with our selective excitation method. One of them is the enhancement of the single photon generation rate by fabricating a cavity structure on the fiber, owing to the Purcell effect~\cite{purcell01}. This can be realized by coupling the ion and the optical fiber cavity structure such as fiber Bragg grating ~\cite{cavity03} or 1-D photonic crystal cavity ~\cite{photocrys01}. RE ions intrinsically have a low spontaneous emission rate compared to other artificial single photon emitters, typically in the order of kilohertz, and thus increasing the emission rate would be crucial for further applications. In this work, we measure the optical lifetime of the single Nd$^{3+}$ ion which clarifies this point.

Another important feature of our selective excitation method is that in principle, one can optically operate multiple isolated ions within the same fiber individually. This feature is not present in our previous scheme (Figure \ref{fig.selective}(a)) as this method excites the ions simultaneously with a single pump beam. If a cavity structure is fabricated in the fiber~\cite{cavity02, cavity03, photocrys01}, one can induce interaction between the ions mediated by the cavity photons. This potentially constructs a multi-qubit processing unit~\cite{cavity01}. Furthermore, a protocol to entangle the internal energy states of two RE ions, which encode qubits, through the detection of single photons emitted from these ions can be considered. The principle of that method has been experimentally achieved by integrating the ions into cavity-structured crystalline devices and fiber cables~\cite{entanglement01}. We note that for such applications, the coherence time and indistinguishability of the emitted photons should be characterized. In this paper we focus on the single photon generation at room temperature, although the characterization of the coherence time would clarify the necessity of cryogenic environment.

From the features mentioned above, we consider our system as a promising component for all-fiber-integrated quantum networks. Our system is compact, low cost, and straightforward to integrate into a fiber-based communication network as the device is fabricated from a commercially available optical fiber. Importantly, telecom wavelength photons can be generated by selecting specific RE ions and their transitions~\cite{communication01}.

\section{method}

 \begin{figure}[htp]
 \centering
\includegraphics[width=8cm]{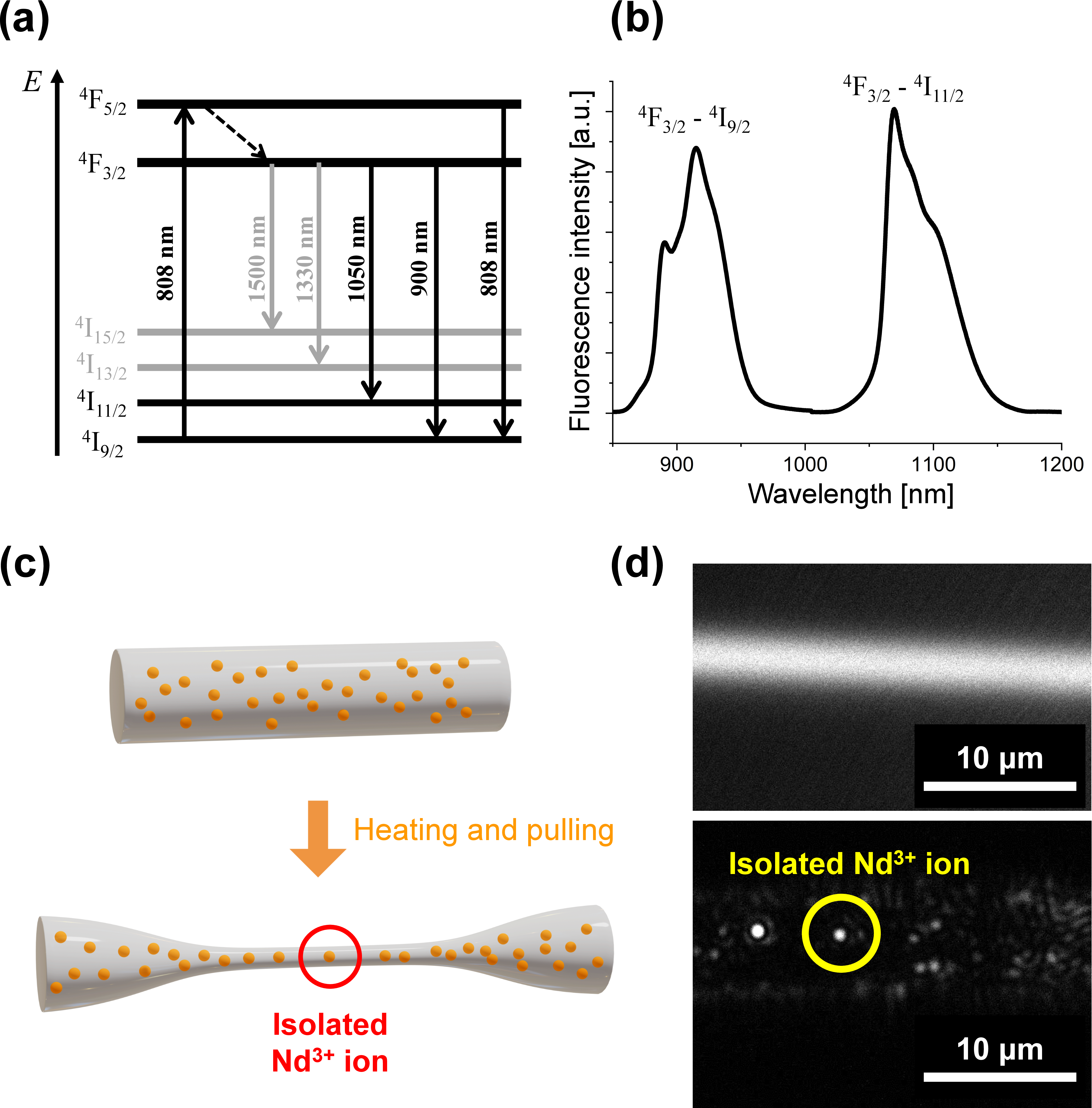}
\caption{\label{fig.concept}
(a) Typical energy level structure of the Nd$^{3+}$ ion in the silica fiber.
(b) Emission spectrum from the Nd$^{3+}$ ions in the unstructured fiber which we used in the experiment. The spectrometer used in the mesurement of this spectrum was intensity calibrated at each wavelength.
(c) Schematic picture of the procedure to obtain spatially isolated Nd$^{3+}$ ions by the tapering process of the fiber.
(d) The photoluminescence images of Nd$^{3+}$ ions in the unstructured fiber(top) and in the tapered fiber(bottom), which were excited by a CW laser($808\,\mathrm{nm}$).
}
 \end{figure}
 
In our experiment, neodymium ion(Nd$^{3+}$) was used as the dopant of the optical fiber. Figure \ref{fig.concept}(a) shows the typical energy level structure of the Nd$^{3+}$ ion in the silica fiber~\cite{Ndspectrum01}. As one can see from Figure \ref{fig.concept}(a), the rich energy level structure of the Nd$^{3+}$ ion allows us to select the wavelength of the generated single photons by using a wavelength filter.
Figure \ref{fig.concept}(b) shows the emission spectrum of the Nd$^{3+}$ ions doped in the unstructured silica fiber which we used in the experiment. This emission spectrum was measured by exciting the ions with a 808 nm pump laser, and using a Czenry-Turner type spectrometer(HRS-300, Princeton Instrument, resolution: 300 gratings/mm). The shape of the emission spectrum suggests a slight influence of the Stark effect by the host material~\cite{Stark01}.

Figure \ref{fig.concept}(c) illustrates the procedure to fabricate the tapered Nd$^{3+}$ ions-doped fiber. The spatially isolated Nd$^{3+}$ ions were obtained by heating and pulling a commercially available single mode fiber of which Nd$^{3+}$ ions doped in the core~\cite{tapering01}. Our single photon source using a single Nd$^{3+}$ ion doped in the fiber does not require cryogenic systems, since the RE ions in the silica fiber are optically active at room temperature. These features make our system a low-cost and compact single-photon source~\cite{Yb01, Nd01}.

Figure \ref{fig.concept}(d) shows the photoluminescence images from the Nd$^{3+}$ ions in the unstructured fiber(top) and in the tapered fiber(bottom), respectively. Both images were taken by a CMOS camera.
The measurement of the optical lifetime, or the autocorrelation measurement were held on the emission from the isolated Nd$^{3+}$ ion marked with a circle in Figure \ref{fig.concept}(d).

\begin{figure}[htp]
 \centering
\includegraphics[width=8cm]{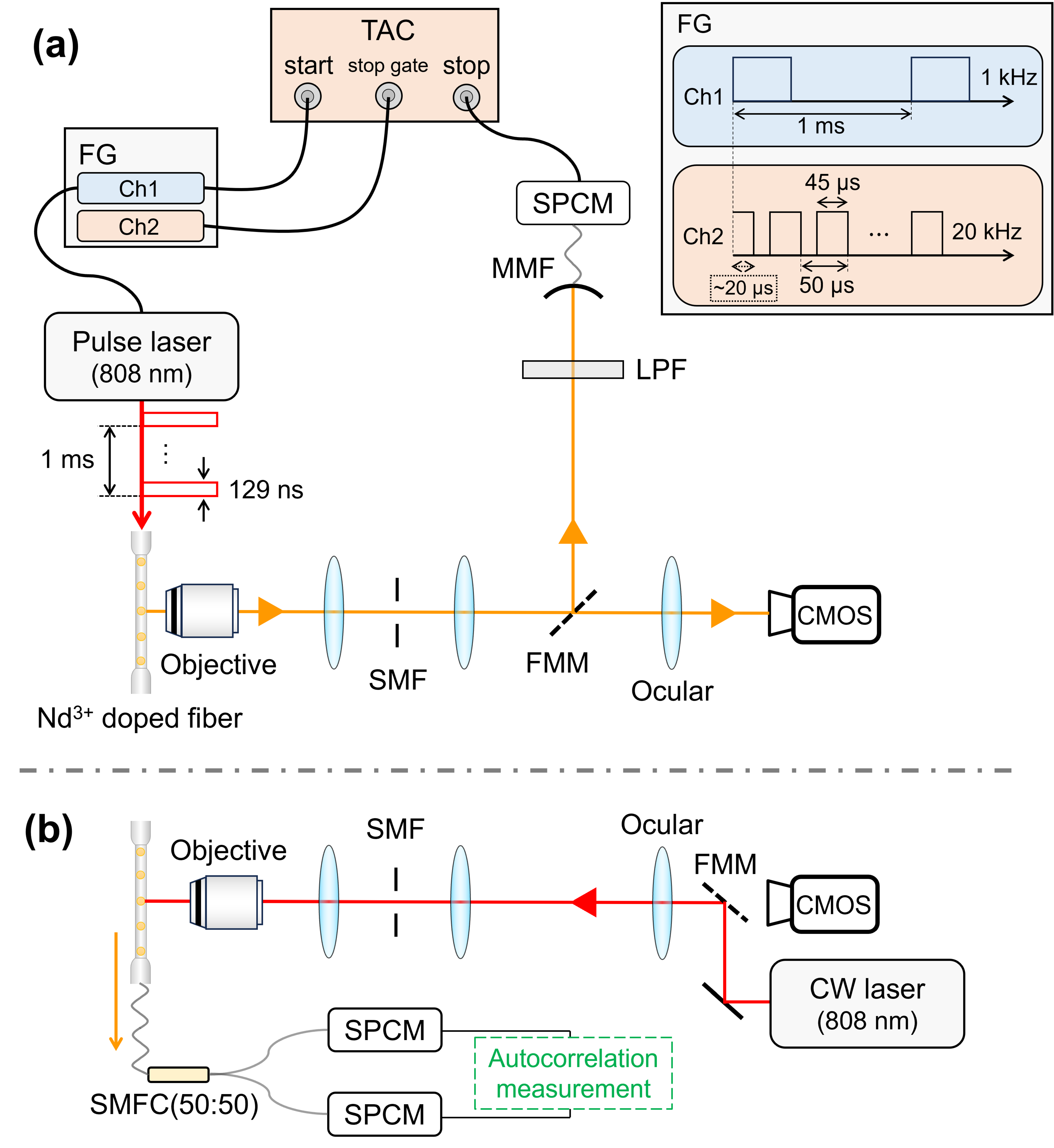}
\caption{\label{fig.method}
(a) Experimental setup for the measurement of optical lifetime.
A pulse laser(wavelength: 808 nm, pulse width: 129 ns) triggered by the function generator(FG) was used as a pump laser. Photons emitted from the Nd$^{3+}$ ion were collected by an objective lens from the side of the fiber. The optical path was changed by the flip mirror mount(FMM). Emitted photons were collected by a multi-mode fiber(MMF) and detected by a single-photon counting module(SPCM) after passing through a longpass filter(LPF, cut-on wacelength: 850 nm). The fluorescence intensity from the Nd$^{3+}$ ion was measured by a time-to-amplitude converter(TAC).
Inset: the FG, which has dual output channels as Ch1 and Ch2, sends the pulses to the pump laser and the TAC.
(b) Experimental setup for the autocorrelation measurement by the selective excitation of a single Nd$^{3+}$ ion in the tapered fiber. Emitted photons were separated into two paths by a 50:50 single-mode fiber coupler(SMFC, $808\pm15\,\mathrm{nm}$), and detected by SPCMs. 
}
  \end{figure}

Figure \ref{fig.method}(a) shows our experimental setup to perform the measurement of the optical lifetime.
To perform the measurements on a single Nd$^{3+}$ ion, we have first observed the photoluminescence from the Nd$^{3+}$ ions by aligning the CMOS camera at the side of the fiber. We have selected one of the ions (i.e. one of the spots observed with the camera) and inserted a spatial mode filter(SMF) to isolate it from the rest of the ions.
The SMF consists of two lenses and an iris(hole size: $1.2\,\mathrm{mm}$).
When measuring the optical lifetime, the ion was pumped by injecting a pulse laser (wavelength: 808 nm, pulse width: 129 ns) into the guided mode of the fiber.
The pulse laser was triggered by a function generator(FG), which has dual output channels which we labeled them as Ch1 and Ch2, respectively. The trigger pulse was sent to the pulse laser from Ch1(repetition rate: $1\,\mathrm{kHz}$).
Photons emitted from the isolated Nd$^{3+}$ ion were collected by an objective lens($\mathrm{N.A.}=0.5$, working distance: $10.6\,\mathrm{mm}$) placed at the side of the fiber.
The optical path was changed by a flip mirror mount(FMM), and the emitted photons were collected by a multi-mode fiber(MMF) connected to a single-photon counting module(SPCM, PerkinElmer Inc., SPCM-AQR-14-FC) after passing through a longpass filer(LPF, Thorlabs Inc., FELH0850), whose cut-on wavelength is $850\,\mathrm{nm}$ and the optical density is more than $5$ at the wavelength of the pump laser.

The fluorescence intensity from a single Nd$^{3+}$ ion was measured by a SPCM and a time-to-amplitude converter(TAC) to detect weak emission signals from the ion. The start signal was sent from Ch1 of the FG to the TAC at the identical time when the same trigger pulse was sent to the pulse laser.
Our measurement time range was $500\,\mathrm{\mu s}$, the emission rate was about $\sim 1\,\mathrm{kHz}$, and the dark count rate was about $\sim 0.5\,\mathrm{kHz}$. Under this condition, it is difficult to perform the measurement with low uncorrelated signals since the uncorrelated stop signals are detected during the time range of the inverse of the emission rate.
Therefore, to exclude the uncorrelated signals by shortening the active measurement time, the TAC was set not to detect stop signals while a gate pulse was sent to the stop gate from Ch2(repetition rate: $20\,\mathrm{kHz}$, width: $45\,\mathrm{\mu s}$) of the FG as shown in Figure \ref{fig.method}(a). Under this condition, the total active measurement time was $50\,\mathrm{\mu s}$($5\,\mathrm{\mu s} \times 10$).
We note that the scattered pump photons do not practically affect the results of the optical lifetime measurement. This is because that none of the incoming photons were counted during the time width of the gate pulse. The time width of the first gate pulse after the pump pulse injection is $\sim 20\,\mathrm{\mu s}$. This time is much longer than the pulse width of the pump laser, which is 129 ns.

We used a silicon avalanche photodiode as the SPCM in the setup. The quantum efficiencies of the SPCM are $\sim 35\,\%$ and $\sim 5\,\%$ at the wavelength of $900\,\mathrm{nm}$ and $1050\,\mathrm{nm}$, respectively.
Judging from the spectrum in Figure \ref{fig.concept}(b), the fluorescence intensity from the transitions of $900\,\mathrm{nm}$ and $1050\,\mathrm{nm}$ are comparable. This indicates that the ratio of the photons measured from these two transitions is mainly determined by the quantum efficiency of the SPCM, which is about $7:1$. Therefore, we consider that the main contribution in our optical lifetime measurement is the photons from the 900 nm transition. On the other hand, photons emitted from the $1050\,\mathrm{nm}$ transition are not excluded completely. In principle, individual transitions should be distinguished by selecting each bandwidth of the transition using narrow band filters to evaluate the indistinguishability of emitted single photons. However, observing photons emitted from individual transitions is technically hard due to the low emission rate of the Nd$^{3+}$ ion with the current setup.

In the measurement of the optical lifetime, we used the method for exciting the Nd$^{3+}$ ions simultaneously with the pump laser injected from the direction of the guided mode of the fiber, which we studied in our previous work~\cite{Yb01, Nd01}. We refer to this as the non-selective excitation hereafter.
In contrast, the autocorrelation measurement was performed by our newly developed method where only one isolated Nd$^{3+}$ ion within the tapered fiber was excited selectively by injecting the pump laser(CW, 808 nm) from the side of the fiber, as shown in Figure \ref{fig.method}(b). We refer to this method as the selective excitation. Photons emitted from the excited single Nd$^{3+}$ ion were guided though the tapered fiber, and then coupled to a 50:50 single mode fiber coupler(SMFC, $808\pm15\,\mathrm{nm}$, Thorlabs, Inc., TN808R5F1).
These photons were separated into two paths by the SMFC, and detected by two SPCMs.
Using a laser source, we have confirmed that the split ratio and the transmission ratio of the SMFC around 808 nm to be $1:1$ and 80\%, respectively. On the other hand, photons outside of the target wavelength range is not excluded, as we measure the split ratio of $1:3$ and the transmission ratio of 40\% around the wavelength of 900 nm.
However, the autocorrelation measurement was conducted under the condition where the single count rates on both SPCMs are balanced to $1:1$. Therefore, we regard the detected photons in the autocorrelation measurement by the selective excitation were mainly generated by the fluorescence under near resonant condition.

\section{result}

\subsection{lifetime}
\begin{figure}[htp]
 \centering
\includegraphics[width=8cm]{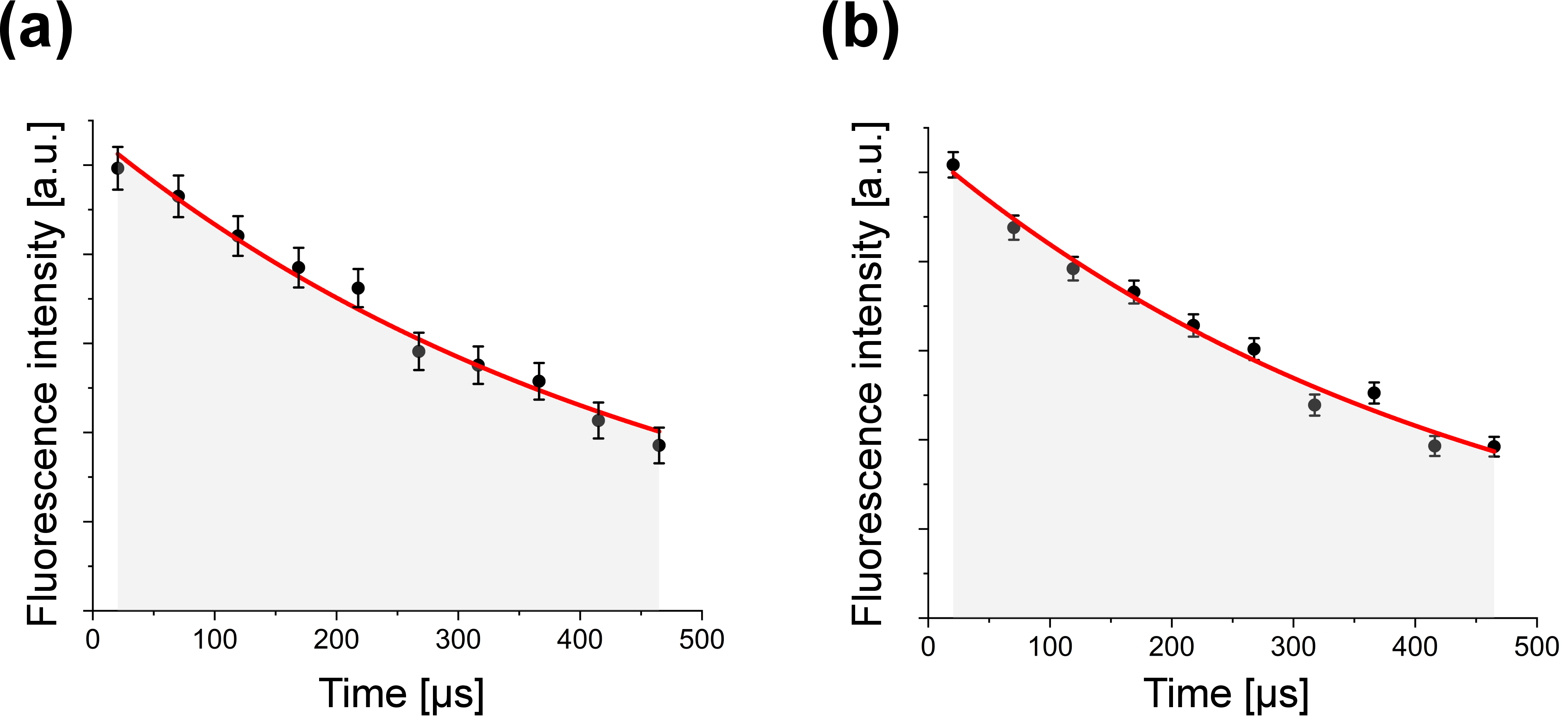}
\caption{\label{fig.result}
Measurement result of the measurement of optical lifetime with a fitted curve(red solid curve) and estimated error bars (a) for an ensemble of Nd$^{3+}$ ions in the unstructured silica fiber and (b) for a single Nd$^{3+}$ ion in the tapered silica fiber.
}
  \end{figure}

We have measured the optical lifetime of a single neodymium ion in the tapered fiber to estimate its influence on the single-photon correlation time of the second-order autocorrelation function described in the following subsection.
The relationship between the fluorescence intensity and the delay time is expressed as~\cite{lifeeq01}
  \begin{equation}
I(t)=I_{0} \exp \left( -\dfrac{t}{\tau_{\mathrm{life.}}}\right), \label{eq.lifetime}
  \end{equation}
where $I(t)$ is the fluorescence intensity, $t$ is the delay time, $\tau_{\mathrm{life.}}$ is the optical lifetime and $I_{0}$ is the fluorescence intensity at $t=0$.
Figure \ref{fig.result}(a) shows the measurement result of the optical lifetime of an ensemble of Nd$^{3+}$ ions in the unstructured silica fiber, which was measured by the same setup as shown in Figure \ref{fig.method}(a). Fitting this result with Eq. (\ref{eq.lifetime}), the optical lifetime of the ensemble of Nd$^{3+}$ ions was estimated as $475\pm22\,\mathrm{\mu s}$.
Figure \ref{fig.result}(b) shows the result when measured from a single Nd$^{3+}$ ion in the tapered silica fiber. Fitting this result with Eq. (\ref{eq.lifetime}), the optical lifetime of a single Nd$^{3+}$ ion was estimated as $452 \pm 22 \,\mathrm{\mu s}$.
The experimentally obtained values for both the ensemble or a single Nd$^{3+}$ ion resulted to be close within the error margin.
Therefore, we can state that the tapering process of the fiber do not affectively change the optical lifetime.

\subsection{autocorrelation}
The autocorrelation measurements were performed with both the selective and the non-selective excitation method to obtain the evidence that the spot marked in Figure \ref{fig.concept}(d) contained only a single Nd$^{3+}$ ion.
The second-order autocorrelation function is given as~\cite{g2.eq01}
  \begin{align}
g^{(2)}(\tau) = 1- \left\{ 1-g^{(2)}(0)  \right\} \exp \left[ -\left( \dfrac{1}{\tau_{\mathrm{abs.}}} + \dfrac{1}{\tau_{\mathrm{life.}}} \right)\tau \right], \label{eq.g2}
  \end{align}
where $\tau_{\mathrm{abs.}}$ is the absorption time, $\tau_{\mathrm{life.}}$ is the optical lifetime, $\tau$ is the delay time, and $g^{(2)}(0)$ is the value of the second-order autocorrelation function at zero-delay.
The optical lifetime of the Nd$^{3+}$ ion is a few hundred microseconds order, which we measured in the previous section. In general, this value is sufficiently longer than that of the absorption time.
Therefore, under such condition, Eq. (\ref{eq.g2}) can be expressed as $g^{(2)}(\tau) \simeq 1- \left\{ 1-g^{(2)}(0)  \right\} \exp \left(-\tau/\tau_{\mathrm{abs.}}\right)$, which indicates that the correlation time of the second-order autocorrelation function is practically determined by the absorption time $\tau_{\mathrm{abs.}}$.

  \begin{figure}[htp]
 \centering
\includegraphics[width=8cm]{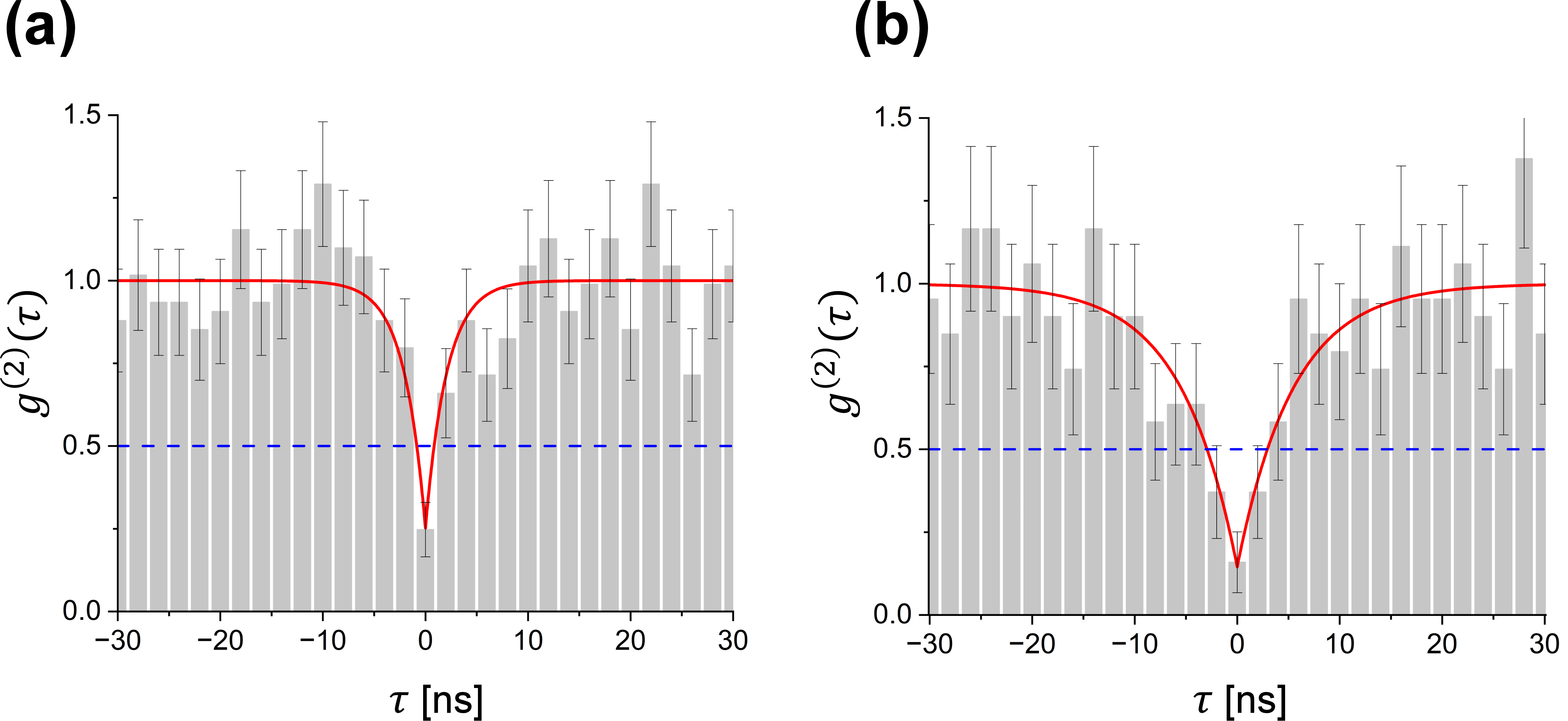}
\caption{\label{fig.result2}
Result of the autocorrelation measurement on the single Nd$^{3+}$ ion with a fitted curve(red solid curve) and estimated error bars. (a) Was done by the non-selective excitation and (b) was done by the selective excitation method.
}
  \end{figure}

Figure \ref{fig.result2}(a) shows the results of the autocorrelation measurement by the non-selective excitation method. From the fitted curve, the value of the second-order autocorrelation function at zero-delay was estimated as $g^{(2)}(0)= 0.25 \pm 0.14$, and the correlation time was estimated as $2\tau_{\mathrm{abs.}}=4.2 \pm 1.2 \,\mathrm{ns}$.
Figure \ref{fig.result2}(b) shows the result of the autocorrelation measurement by the selective excitation method.
From the fitted curve, the value of second-order autocorrelation function at zero-delay was estimated as $g^{(2)}(0)= 0.15 \pm 0.12$, and the correlation time was estimated as $2\tau_{\mathrm{abs.}}=11.0 \pm 3.1 \,\mathrm{ns}$. In both methods, the experimental value of $g^{(2)}(0)$ being less than $0.5$ is the evidence of single photon generation. This also indicates that the result of autocorrelation measurement with the selective excitation method was not largely affected by the absorption and emission process of other doped Nd$^{3+}$ ions within the measurement time range.
One of the possible reasons that may have raised the value of $g^{(2)}(0)$ is the uncorrelated background noise from the pump light due to the fiber imperfections. In particular, in the selective excitaion method, some of the scattered pump light may be coupled into the fiber guided modes due to the fiber imperfections created by the tapering process~\cite{hbn01}.

\section{Discussion}
Here, we compare the conditions and results of the selective excitation and the non-selective excitation methods.
In our experiment, the correlation time of the second-order autocorrelation function is determined by the absorption time. Under the resonant condition, the absorption time is represented as
  \begin{equation}
\tau_{\mathrm{abs.}} = \dfrac{1}{BU}, \label{eq.abs01}
  \end{equation}
where $B$ is the Einstein coefficient and $U$ is the energy density per unit frequency of the pump light.
The intensity of the pump light $I_{\mathrm{pump}}$ is represented as~\cite{Nd01}
  \begin{equation}
I_{\mathrm{pump}} \simeq  \dfrac{c^{2} U}{\lambda^{2}}\Delta \lambda, \label{eq.intensity01}
  \end{equation}
where $c$ is the speed of light, $\lambda$ is the wavelength and $\Delta \lambda$ is the broadening of the wavelength of the pump light.
From Eq. (\ref{eq.abs01}), (\ref{eq.intensity01}) and the Planck's law $A/B=8\pi h/\lambda^{3}$, where $A$ is the Einstein coefficient and $h$ is the Planck's constant, the relationship between the absorption time and the intensity of the pump light is represented as
  \begin{equation}
\tau_{\mathrm{abs.}} = \dfrac{8\pi hc^{2}\Delta \lambda \tau_{\mathrm{life.}}}{\lambda^{5}}\times \dfrac{1}{I_{\mathrm{pump}}}, \label{eq.abs02}
  \end{equation}
where $\tau_{\mathrm{life.}}=1/A$ is the optical lifetime. Therefore, the absorption lifetime is inversely proportional to the intensity of the pump light as $\tau_{\mathrm{life.}} \propto I_{\mathrm{pump}}^{-1}$.
In the selective excitation method, the pump power was $\sim 2.0\,\mathrm{\mu W}$, which was measured before incident on the objective lens. The diameter of the pump light after focused by the objective lens was about $\sim 9\,\mathrm{\mu m}$, which was calculated from the hole size of the iris and the working distance of the objective lens based on gaussian optics. Therefore, the intensity of the pump light was calculated as $\sim 3\times 10^{4} \,\mathrm{W/m^{2}}$.
On the other hand, in the non-selective excitation method, the power of the pump light propagating in the fiber was $\sim 0.2\,\mathrm{\mu W}$.
The tapered fiber used here was fabricated with the same method in our previous work~\cite{Yb01}. Therefore, the diameter of the Nd$^{3+}$ ions-doped tapered fiber was estimated to be comparable to the value measured then by a scanning electron microscope, which was about $\sim 2\,\mathrm{\mu m}$.
From these parameters, the intensity of the pump light in the non-selective excitation method was calculated as about $\sim 6 \times 10^{4}\,\mathrm{W/m^{2}}$.
Therefore, the ratio of the pump intensity of the non-selective excitation to the selective excitation was calculated as about $\sim 2$. 
From the results of the autocorrelation measurement(Figure \ref{fig.result2}(a) and \ref{fig.result2}(b)), the ratio of the correlation time measured by the selective excitation to the non-selective excitation was calculated as $2.6 \pm 1.1$.
The ratio given by the excitation intensity approximately agrees with the ratio of the correlation time of the second-order autocorrelation function, which complements our assumption that the correlation time is determined by the absorption time.
The results are summarized in Table \ref{tab1}.

\begin{table*}[htb]
    \begin{center}
    \caption{Experimental parameters for calculating the pump intensity in the selective excitation and the non-selective excitation methods.}
    \label{tab1}
    \begin{tabular}{ccc}
        \hline
         & Selective excitation & Non-selective excitation \\ \hline
        Pump power $[\mathrm{\mu W}]$ & $\sim 2.0$ & $\sim 0.2$ \\ 
        Diameter of the pump light $[\mathrm{\mu m}]$ & $\sim 9$ & $\sim 2$ \\ 
        Pump intensity $\mathrm{[\times 10^{4}\,W/m^{2}]}$ & $\sim 3$ & $\sim 6$  \\ \hline
        Correlation time  [ns] &  $11.0\pm3.1$  & $4.2\pm1.2$   \\ \hline
    \end{tabular}
\end{center}
\end{table*}

Secondly, we compare the experimentally obtained collection efficiencies of the selective and non-selective excitation methods based on the results of the autocorrelation measurement.
The collection efficiency of the selective excitation $\eta_{\mathrm{s}}$ is defined as the proportion of the detected photon count rate $N_{\mathrm{s}}$ to the actual emission rate from a single Nd$^{3+}$ ion $n_{\mathrm{s}}$, which is represented as $\eta_{\mathrm{s}} = N_{\mathrm{s}} / n_{\mathrm{s}}$.
The collection efficiency of the non-selective excitation $\eta_{\mathrm{f}}$ is also defined similary as $\eta_{\mathrm{f}} = N_{\mathrm{f}} / n_{\mathrm{f}}$.
The ratio of the two collection efficiencies of the selective excitation method to the non-selective excitation method can be represented as
\begin{equation}
\dfrac{\eta_{\mathrm{s}}}{\eta_{\mathrm{f}}} = \dfrac{N_{\mathrm{s}}}{N_{\mathrm{f}}} \cdot \dfrac{n_{\mathrm{f}}}{n_{\mathrm{s}}},
\end{equation}
The values of each single count rate are $N_{\mathrm{s}}=585\pm13\,\mathrm{Hz}$ and $N_{\mathrm{f}}=927\pm12\,\mathrm{Hz}$. These values are calculated by the coincidences in the autocorrelation measurement by each excitation method~\cite{scirep01}, which in principle contain the uncorrelated background noise from the pump light.
Assuming that the actual emission rate is proportional to the pump intensity, $n_{\mathrm{f}}/n_{\mathrm{s}}$ corresponds to the ratio of the correlation time of the second-order autocorrelation function given by the values in Table \ref{tab1}. This leads to $n_{\mathrm{f}}/n_{\mathrm{s}} \sim 2.6.$
Hence we obtain $\eta_{\mathrm{s}} / \eta_{\mathrm{f}} \sim 1.6$, meaning that the experimental collection efficiency of the selective excitation method is roughly $1.6$ times larger than that of the non-selective excitation method.

Although the calculation shows the greater collection efficiency of the selective excitation method, it is still limited by the number of the propagation modes in the tapered fiber. This is due to the taper fiber satisfying the multi-mode condition since the air plays a role of the cladding layer. A straightforward way to improve the collection efficiency with our current device is to collect the photons from both directions of the tapered fiber, instead of single direction as we performed in Figure \ref{fig.method}(b).

\section{Conclusion}
In conclusion, we experimentally demonstrated the generation of single photons at room temperature by selectively exciting an isolated single RE ion within an optical tapered fiber. In our method, solely a single RE ion inside the tapered silica fiber is selectively excited by injecting the pump laser from the side of the fiber, and photons emitted from the excited single RE ion could be efficiently collected by channeling them through the tapered fiber.
The value of the second-order autocorrelation function at zero-delay was obtained as $g^{(2)}(0) = 0.15 \pm 0.12$ by the selective excitation. This value of $g^{(2)}(0)$ being less than $0.5$ is the evidence of single-photon generation.
In addition, we measured the optical lifetime of a single Nd$^{3+}$ ion in the tapered silica fiber. The obtained result was $\tau_{\mathrm{life.}}=452\pm22\,\mathrm{\mu s}$, which is comparable to that of an ensemble of Nd$^{3+}$ ions in the unstructured silica fiber. Therefore, it is expected that the tapering process does not affectively change the optical lifetime of the RE ion.
The long optical lifetime indicates that the correlation time of the second-order autocorrelation function is practically determined by the absorption time, instead of the optical lifetime, given that the measurement time range is much smaller than the optical lifetime.
This was confirmed by estimating the ratio of the pump intensity of our two excitation methods which agreed with the ratio of their correlation time.
As our emitter utilizes a commercially available RE-doped fiber while achieving the efficient coupling of the emitted photons to the fiber guided mode at room temperature, it can become a low-cost key component for various photonic quantum technologies including all-fiber-integrated quantum communication networks.


\appendix

\section{Collection efficiency of emitted photons among different methods}

\begin{figure}[htp]
 \centering
\includegraphics[width=9cm]{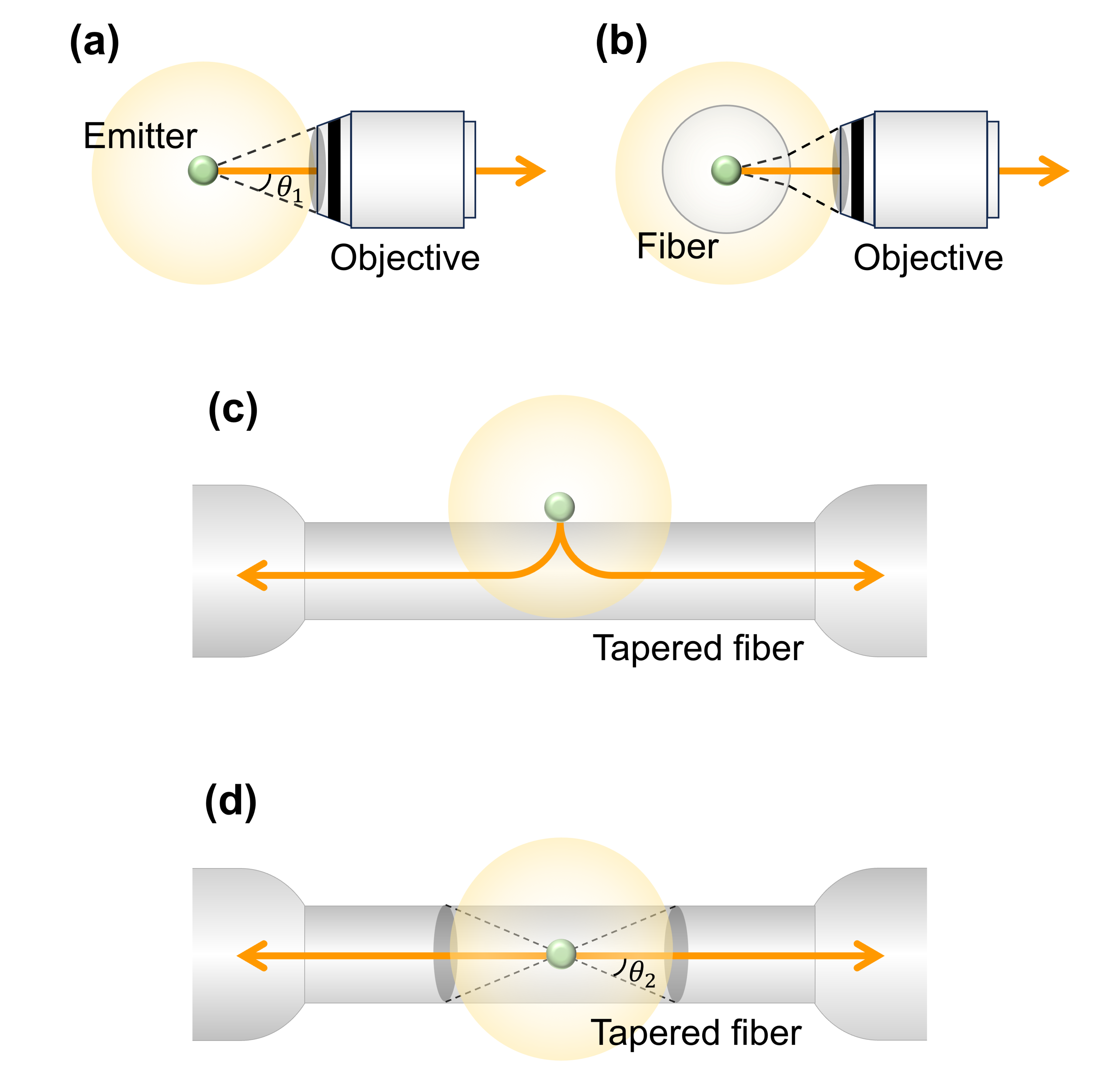}
\caption{\label{fig.channeling}
The comparison of different methods to collect photons from an emitter.
(a) Collecting the photons from an emitter placed in free space with an objective lens.
(b) Collecting the photons from an emitter doped in the fiber with an objective lens.
(c) Coupling a tapered fiber to an emitter positioned at the vicinity of the fiber.
(d) Collecting the photons from an emitter inside a tapered fiber by channeling them through the guided mode of the fiber.
}
 \end{figure}

In this Appendix, we calculate and compare the collection efficiencies of the photons among different experimental methods which combine a single photon emitter and the tapered fiber. 
Figure \ref{fig.channeling}(a) shows the method to collect the emitted photons using an objective lens.
In this method, the proportion of the photons that can be collected is determined by the solid angle from the emitter to the objective lens. The solid angle $\Omega_{1}$ is given by $\Omega_{1} = 2\pi(1-\cos \theta_{1})$, where $\theta_{1}$ is the maximum angle of the photons that can be collected, which is related to the numerical aperture of the objective lens. The value of the numerical aperture(N.A.) is given by $\mathrm{N.A.} = n \sin \theta_{1}$, where $n$ is the refractive index of the medium. Therefore, the collection efficiency $\eta_{1}$ is represented as
  \begin{equation}
\eta_{1} = \dfrac{\Omega_{1}}{4\pi} = \dfrac{1}{2} \left\{ 1 - \sqrt{1 - \left( \dfrac{\mathrm{N.A.}}{n} \right)^2 } \right\}. \label{eq.objective}
  \end{equation}
The value of the numerical aperture of the objective lens in our experiment was $\mathrm{N.A.} = 0.5$. Thus from Eq. (\ref{eq.objective}), the collection efficiency is calculated to be about $\sim 6.7\,\%$, where we regarded as $n \sim 1$. From Eq. (\ref{eq.objective}), the collection efficiency can be increased by using an objective lens with greater N.A., which can reach up to $50\,\%$ at $\mathrm{N.A.} \rightarrow 1$. This value is the theoretical limit for the method using the objective lens.

When we take into account the existence of the tapered fiber enclosing the ion, as shown in Figure \ref{fig.channeling}(b), the proportion of photons that can reach the objective lens decreases. This is mainly due to the refraction and reflection of the photons at the fiber-to-air boundary. FDTD simulations suggest that the collection efficiency is up to $3\sim5\%$ depending on the position of the ion within the fiber. Here we used the same N.A. of $0.5$, the fiber refractive index of $1.45$, and the fiber diameter of $2 \,\mu\mathrm{m}$. 

A well studied method to couple an emitter to a fiber mode is to place the emitter in the vicinity of the tapered fiber, as illustrated in Figure \ref{fig.channeling}(c)~\cite{qdot01, qdot02, qdot03, molecule01, hbn01, nanodiamond01}. In this method, the maximum theoretical value of the channeling efficiency is estimated to be about $30\,\%$~\cite{coupling01}. However, the maximum channeling efficiency is realized by optimizing the conditions such as the distance between the emitter and the tapered fiber, and the core diameter of the fiber. To the best of our knowledge, the experimentally achieved channeling efficiency with the method was at maximum $\sim 20\,\%$~\cite{coupling01, coupling02}.

In comparison, our method utilizes the RE ions initially doped within the fiber as the emitter, as illustrated in Figure \ref{fig.channeling}(d). Therefore, the channeling efficiency $\eta_{2}$ is determined regardless of the conditions mentioned above.
The maximum radiation angle $\theta_{2}$ which causes the total reflection on the surface of the fiber whose refractive index is $n'$ is given by $\sin \left(\pi/2 - \theta_{2} \right) = \cos \theta_{2} = 1/n'$ from the geometrical relation of $\theta_{2}$. The channeling efficiency $\eta_{2}$ is given by the solid angle $\Omega_{2}=2\pi \left( 1-\cos \theta_{2} \right)$ from the RE ion to the fiber as~\cite{Yb01}
  \begin{equation}
 \eta_{2} = \dfrac{\Omega_{2}}{4\pi}  = \dfrac{1}{2} \left( 1 - \dfrac{1}{n'} \right). \label{eq.channeling}
  \end{equation}
The refractive index of the silica fiber for a wavelength around $808\,\mathrm{nm}$ is $n' \sim 1.45$.
From Eq. (\ref{eq.channeling}), the maximum channeling efficiency when the photons are collected from both directions of the fiber is calculated as about $31\,\%$.
This value of channeling efficiency can be achieved independent of the position of the emitter or the core diameter of the fiber within the scope of the geometrical optics.
It was observed through FDTD simulation that the channeling efficiency was increased up to about $\sim 38\,\%$ when the emitter was displaced from the center of the fiber~\cite{Yb01}.



\section*{Acknowledgements}
The authors would like to thank Mark Sadgrove for supporting to prepare the tapered fiber samples and for helpful discussions.
The authors would also like to thank Shigeki Takeuchi, Ryo Okamoto, Yu Mukai and Hideaki Takashima for helpful discussions.
Kaito Shimizu has been supported by a doctoral student scholarship(2023) of Amano Institute of Technology, Japan.




\end{document}